\documentclass{emulateapj}
\usepackage{multirow}
\def\msun{$M_{\odot}$}

\def\ergcm{\hbox{erg cm$^{-2}$ s$^{-1}$ }} 
 
\def\xte{{\it RXTE~}}

\shortauthors{Lin et al.}
\begin{document}

\title{{\it Suzaku} and {\it BeppoSAX} X-ray Spectra of the Persistently Accreting Neutron-Star Binary 4U 1705-44}

\author{Dacheng Lin\altaffilmark{1,2}, Ronald A. Remillard\altaffilmark{1}, and Jeroen Homan\altaffilmark{1}}
\affil{$^1$ MIT Kavli Institute for Astrophysics and Space Research, MIT, 70 Vassar Street, Cambridge, MA 02139-4307, USA}
\affil{$^2$ Centre d'Etude Spatiale des Rayonnements, 9, av du Colonel Roche, BP 44346, 31028 Toulouse Cedex 4, France, email: Dacheng.Lin@cesr.fr}

\begin{abstract}
We present an analysis of the broad-band spectra of 4U~1705--44
obtained with {\it Suzaku} in 2006--2008 and by {\it BeppoSAX} in
2000. The source exhibits two distinct states: the hard state shows
emission from 1 to 150 keV, while the soft state is mostly confined to
be $<40$ keV. We model soft-state continuum spectra with two thermal
components, one of which is a multicolor accretion disk and the other
is a single-temperature blackbody to describe the boundary layer, with
additional weak Comptonization represented by either a simple power
law or the SIMPL model by Steiner et al. The hard-state continuum
spectra are modeled by a single-temperature blackbody for the boundary
layer plus strong Comptonization, modeled by a cutoff power law. While
we are unable to draw firm conclusions about the physical properties
of the disk in the hard state, the accretion disk in the soft state
appears to approximately follow $L\propto T^{3.2}$. The deviation from
$L\propto T^4$, as expected from a constant inner disk radius, might
be caused by a luminosity-dependent spectral hardening factor and/or
real changes of the inner disk radius in some part of the soft
state. The boundary layer apparent emission area is roughly constant
from the hard to the soft states, with a value of about 1/11 of the
neutron star surface. The magnetic field on the surface of the NS in
4U~1705--44 is estimated to be less than about $1.9\times 10^8$ G,
assuming that the disk is truncated by the ISCO or by the neutron star
surface.  Broad relativistic Fe lines are detected in most spectra and
are modeled with the diskline model. The strength of the Fe lines is
found to correlate well with the boundary layer emission in the soft
state. In the hard state, the Fe lines are probably due to illumination
of the accretion disk by the strong Comptonization emission.

\end{abstract}


\section{INTRODUCTION}
\label{sec:intro}
There are two main classes of luminous and weakly magnetized neutron
stars (NSs) in low-mass X-ray binaries (LMXBs), i.e., atoll and Z
sources, named after the patterns that they trace out in X-ray
color-color diagrams (CDs) or hardness-intensity diagrams (HIDs)
\citep{hava,va2006}. Atoll sources have lower luminosities
($\sim$0.001--0.5 $L_{\mathrm{Edd}}$) than Z sources and have two
distinct X-ray states, i.e., hard (energy spectra are roughly flat
with power-law photon index near 1.7) and soft states (energy spectra
follow exponential decrease above $\sim$10 keV). There is also a
``transitional'' state between these two. The hard, transitional and
soft states of atoll sources are also often referred to as the
``extreme island'', ``island'', and ``banana'' states/branches,
respectively. Z sources only have soft spectra, but there are three
distinct branches with different spectral and timing
behaviors. Recently, \citet{lireho2009} and \citet{hoetal2010}
analyzed a transiently accreting NS, \object{XTE J1701-462}, which
changed from Z-source behavior to atoll-source characteristics as the
outburst decayed from near/super Eddington luminosity
($L_{\mathrm{Edd}}$) to almost quiescence. These results confirmed
that the behavior that the differences in the properties of the two
classes are due to their different mass accretion rates.

The spectral modeling of NS LMXBs has been controversial for a long
time \citep[see][for a review]{ba2001}. The continuum of soft-state
spectra in both atoll and Z sources are generally described by
two-component models that include a soft/thermal and a
hard/Comptonized component
\citep[e.g.,][]{baolbo2000,oobagu2001,diiabu2000,iadiro2005}, and
there have been two competing models, often referred to as the {\it
  Eastern} model \citep[after][]{miinna1989} and the {\it Western}
model \citep[after][]{whstpa1988}, with different choices of the
thermal and Comptonized components. In the {\it Eastern} model, the
thermal and Comptonized components are described by a multicolor disk
blackbody (MCD) and a Comptonized blackbody, respectively. In the {\it
  Western} model, on the other hand, the thermal component is a
single-temperature blackbody (BB) from the boundary layer, and there
is Comptonized emission from the disk. In the hard state, the spectra
are dominated by a hard/Comptonized component, but a soft/thermal
component is generally still required \citep{chsw1997, baolbo2000,
  chba2001,gido2002b}. \citet{lireho2007} implemented the commonly
used two-component models for two classical transient atoll sources,
i.e., \object{Aql X-1} and \object{4U 1608-52}, and outlined the
problem of model degeneracy for accreting NSs, not only from the
choices of the thermal components, but also from the detailed
description of Comptonized components (i.e., the scattering corona
geometry, the seed photon temperature, etc.). The physical
interpretation of the spectral evolution of these atoll sources
inferred from these different models also varies
substantially. However, none of the tested models produced results
similar to those of black-hole X-ray binaries, i.e., $L_{\rm X}
\propto T^4$ tracks for the MCD component and weak Comptonization for
the soft-state spectra when the integrated rms variability in the
power density spectrum (0.1-10 Hz) is only a few percent
\citep{remc2006}.

\citet{lireho2007} devised a hybrid model for atoll sources, based on
a detailed study two frequently recurring atoll-type transients. This
model uses a BB to describe the boundary layer plus a broken power law
for the hard state, and two strong thermal components (MCD and BB)
plus a constrained broken power law (when needed) for the soft
state. This choice for the soft state offers a weak-Comptonization
solution that differs from the strong Comptonization solution of the
two-component models that had previously dominated the literature. The
results of the application of the hybrid model can be summarized as
follows: both the MCD and BB evolve approximately as $L_{\rm X}
\propto T^4$, the spectral/timing correlations of these NSs are
aligned with the properties of accreting black holes, and the visible
BB emission area is very small but roughly constant over a wide range
of $L_{\rm X}$ that spans both the hard and soft
states. \citet{lireho2009} applied this X-ray spectral model to
\object{XTE J1701-462}, and similar results were obtained for the
observations when the source displayed atoll-source
behavior. Deviations of the MCD from the $L_{\rm X} \propto T^4$ track
were observed when the source was bright and behaved as a Z source,
with the inner disk showing a luminosity-dependent radius
increase. This was interpreted as an effect of maintaining the local
Eddington limit in the inner disk edge as the mass accretion rate
varies.

The hybrid spectral model is still empirical, especially the modeling
of Comptonization. Moreover, this model has only been applied thus far
to extensive data obtained with the {\it Rossi X-ray Timing Explorer}
(\xte). \xte has two pointing instruments, which cover the energy
range from $\sim$2.5 to 250 keV. However, there is emission from X-ray
binaries below this energy range, as the characteristic temperatures
of the thermal components are normally below 3 keV. Thus, it is
important to test this model using broad-band spectra that extend to
photon energies below the sensitivity range of \xte.

In this paper we investigate the persistently bright atoll source,
\object{4U 1705-44} \citep{hava}, which was observed seven times in
2006-2008 by {\it Suzaku} \citep{mibain2007}. One of the important
features of {\it Suzaku} is its broad energy band (0.2--600 keV). We
also analyzed two observations of \object{4U 1705-44} made with {\it
  BeppoSAX} (0.1--300 keV) in 2000. Both {\it Suzaku} and {\it
  BeppoSAX} additionally provide better energy resolutions than \xte
($\sim 2\%$, $8\%$, $18\%$ at 6 keV (FWHM) respectively), and this
capability can be used to better resolve the broad Fe emission
lines. Broad Fe lines are commonly seen in X-ray binaries and provide
another tool for investigating the accretion flow around compact
objects \citep[e.g.,][]{mi2007,camibh2008,camiho2009}.

Timing studies for \object{4U 1705-44}, including the findings of
kilohertz quasi-periodic oscillations (kHz QPOs), have been carried
out using observations with \xte
\citep{fovaka1998,baol2002,olbagi2003}. Spectral studies of this
source have also been carried out, using different kinds of X-ray
detectors and spectral models
\citep[e.g.,][]{baol2002,ditame2005,fibaub2007,pisadi2007,hokava2009,refayo2009,didaia2009}. Several
of these authors also reported the detection of a broad relativistic
Fe line from this source. In this paper we concentrate on the spectral
properties of this source. We describe our data analysis in
\S\ref{sec:reduction}, where we also present the long-term light curves and
color-color diagrams. We perform detailed spectral modeling in
\S\ref{sec:specmod}, for which we provide our physical interpretations
in \S\ref{sec:interpretation}. Finally we give our summary and
discussion.

\section{OBSERVATIONS AND DATA REDUCTION}
\label{sec:reduction}

\begin{figure*} \epsscale{1.0}  
\plotone{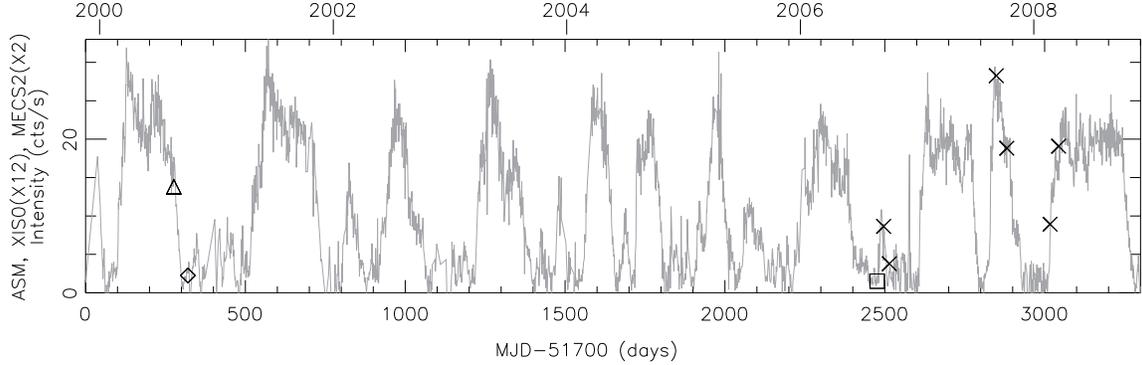}
\caption{The long-term \xte/ASM light curve of 4U~1705--44. The
  various symbols represent different spectral states from different
  pointed observatories: {\it BeppoSAX}/{\it Suzaku} hard state
  (diamond/square) and {\it BeppoSAX}/{\it Suzaku} soft state
  (triangle/cross).
\label{fig:asmdata}} 
\end{figure*}

The long-term light curve of \object{4U 1705-44} is shown in
Figure~\ref{fig:asmdata}. The gray solid line is from the \xte All-Sky
Monitor \citep[ASM;][]{lebrcu1996}, and we can see that the source
displays persistent X-ray emission, with one or two strong intensity
cycles per year. These cycles correspond to state-transition cycles
\citep{hokava2009}. Discrete plot symbols show the time and spectral
states during different pointed observations. The data reduction and
spectral analyses are described below.

\subsection{{\it Suzaku} Data}

\begin{deluxetable*}{lccccccc}
\addtolength{\tabcolsep}{-4pt}
\tabletypesize{\scriptsize}
\tablecaption{The Suzaku observations of 4U~1705--44 in 2006--2008\label{tbl-1}}
\tablewidth{0pt}
\tablehead{Observation ID & \colhead{401046010} & \colhead{401046020} & \colhead{401046030} & \colhead{402051010} & \colhead{402051020} & \colhead{402051030} & \colhead{402051040}\\
  Spectral ID & \colhead{$suz1$} & \colhead{$suz2$} & \colhead{$suz3$} & \colhead{$suz4$} & \colhead{$suz5$} & \colhead{$suz6$} &  \colhead{$suz7$}
}
\startdata
Observation Date & 2006/08/29 & 2006/09/18 & 2006/10/06 & 2007/09/05 & 2007/10/08 & 2008/02/20 & 2008/03/18 \\
Exposure of XIS/PIN (ks)\tablenotemark{a} & 10.9/12.8 & 12.7/12.1 & 13.3/12.4 &  2.2/8.2 & 3.9/14.5 & 18.4/17.3 & 3.0/11.1 \\
XIS Detectors Analyzed & 0 1 2 3 & 0 1 2 3 & 0 1 2 3 & 0 1 3 & 0 1 3 & 0 1 3 & 0 1 3 \\
Window Option & 1/4 & 1/4 & 1/4 & 1/4 & 1/4 & 1/4 & 1/4 \\
Exposure in Burst Option (s) & 1.6 & 2.0 & 2.0 & 0.5 & 0.5 & 2.0 & 0.5\\ 
XIS0 Count Rate (cts/s)\tablenotemark{b} & 18.1 & 103.7 & 45.5 & 339.2 & 225.9 & 107.1 & 229.0 \\
Soft Color\tablenotemark{c} & $0.81\pm0.03$ & $0.72\pm0.02$ & $0.64\pm0.02$ &$0.81\pm0.03$ & $0.73\pm0.02$ & $0.66\pm0.02$ & $0.79\pm0.03$ \\
Hard Color\tablenotemark{c} & $0.34\pm0.03$ & $0.11\pm0.01$ & $0.09\pm0.01$ &$0.08\pm0.01$ & $0.08\pm0.01$ & $0.09\pm0.01$ & $0.09\pm0.01$\\
Radius of Central Region\\
With $>$5\% pile-up (pixels) & 0 & 55 & 32 & 52 & 39 & 60 & 35\\
Radius of Central Region\\
With $>$10\% pile-up (pixels) & 0 & 39 & 21 & 35 & 24 & 42 & 24\\
Spectral state & hard & soft & soft & soft & soft & soft & soft \\
\enddata 
\tablenotetext{a}{The dead time and burst clock options have been
taken into account. All XIS detectors have the same exposure time, and
the values given are for one detector only.}  
\tablenotetext{b}{Total time-averaged count rates, corresponding to the fully integrated PSF.}
\tablenotetext{c}{The average and standard deviation of the colors for each observation based on 128s data bins.}
\end{deluxetable*}

\begin{figure}
\plotone{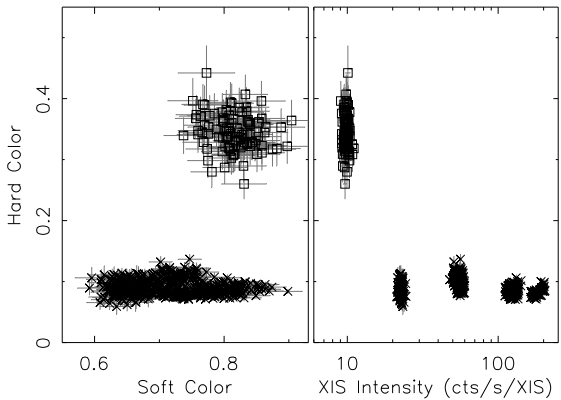}
\caption{Color-color and hardness-intensity diagrams of 4U~1705--44
based on {\it Suzaku} observations in 2006--2008, with bin size 128
s. The squares are from observation 401046010, and the crosses from
the other observations. For the definitions of the soft and hard
colors, see the text. One-$\sigma$ statistical error bars are also
shown.
\label{fig:ccdiag}}
\end{figure}

{\it Suzaku} made seven observations of \object{4U 1705-44} in
2006--2008. Detailed information of these observations is given in
Table~\ref{tbl-1}. Both the X-ray Imaging Spectrometer \citep[0.2--12
  keV, XIS;][]{kotsdo2007} and the Hard X-ray Detector \citep[10--600
  keV, HXD;][]{taaben2007} instruments were used during these
observations. There are four XIS detectors, numbered as 0 to 3. XIS0,
2 and 3 all use front-illuminated CCDs and have very similar responses,
while XIS1 uses a back-illuminated CCD. XIS2 was damaged in 2006
November, and its data are analyzed only for the first three
observations. The HXD instrument includes both PIN diodes (10--70 keV)
and GSO scintillators (30--600 keV). Both the PIN and GSO are
collimated (non-imaging) instruments.

We reprocessed each observation using the \verb|aepipeline| tool
provided by {\it Suzaku} FTOOLS version 15 and applying the latest
calibration available as of 2010 February. We then applied the
publicly available tool \verb|aeattcor.sl| by John E. Davis to obtain
a new attitude file for each observation. This tool corrects the
effects of thermal flexing of the {\it Suzaku} spacecraft and obtains
more accurate estimate of the spacecraft attitude. For all our seven
observations, the above attitude correction produced sharper PSF
images. With the new attitude file, we updated the XIS event files
using the FTOOLS \verb|xiscoord| program.

Since 4U 1705-440 is a relatively bright source, window and burst
options were adopted during each observation to limit the effects of
event pile-up (Table~\ref{tbl-1}). Despite this, pile-up was still
present in the image center. We estimated the pile-up fractions at
different positions of the CCD using the publicly available tool
\verb|pileup_estimate.sl| by Michael A. Nowak. The pile-up fraction
refers to the ratio of events lost via grade or energy migration to
the events expected in the absence of pile-up. The unfiltered pile-up
fractions integrated over the whole CCDs are about 10--15$\%$ for all
observations except for the observation 401046010
($\sim$3$\%$). Table~\ref{tbl-1} lists the radii of the central
circular regions that contain most of the XIS CCD pixels with local
pile-up fractions that exceed 5$\%$ and 10$\%$, respectively. We used
annular regions to extract spectra (circular regions are used for
observation 401046010). The outer radius was set to be 120 pixels,
while two cases of inner radii were used, corresponding to the 5$\%$
and 10$\%$ pile-up exclusion regions, respectively. The corresponding
integrated pile-up fractions of all annular regions are $\sim$3$\%$
and $\sim$5$\%$, respectively. Using the models for the soft-state
spectra, which include MCD and BB components, we find that the
spectral fitting results using 10$\%$ pile-up exclusion regions show
systematic decrease in the soft-component (MCD) flux and increase in
the hard-component (BB) flux, by about 3$\%$, compared with the
results using 5$\%$ pile-up exclusion regions. The differences in most
cases are within the error bars at a 90$\%$ confidence level, and the
conclusions of this paper hold for either case. For simplicity and
increased accuracy, we only show results using the 5$\%$ pile-up
exclusion regions below.

\object{4U 1705-44} is in the direction toward the Galactic ridge, and
the background consists of non-X-ray (particle) background, absorbed
cosmic X-ray background, and Galactic ridge emission. Even though our
source can be regarded as a point source, there is no region of the
detector plane that is free from the source emission to estimate the
pure background. This is because the PSF of the XIS is quite spread
out and our source during these seven observations was bright enough
to dominate over the background (1--10 keV) throughout the $1/4$
window. We compared results using two methods of estimating the
background. In the first method, we set the background region to be
the whole 1/4 window excluding a circular region of radius $350''$
around the source. In the second method, we estimated the non-X-ray
background, which varies with time, using the \verb|xisnxbgen| tool
based on the night Earth data by {\it Suzaku}. We estimated the X-ray
background, including the cosmic X-ray background and Galactic ridge
emission, using observation 100026030 by {\it Suzaku}. It is specific
for observing the background emission around the supernova remnant
\object{RX J1713-3946}, and the pointing direction of this observation
has a $184''$ offset from \object{4U 1705-44}. Both of the methods
turn out to give very similar spectral fit results, and we only show
results using the second method below.

The response files of the XIS for each observation were generated
using the \verb|xisresp| script which uses the \verb|xisrmfgen| and
\verb|xissimarfgen| tools (specifying $1\%$ accuracy). They take into
account the time variation of the energy response and the specific
extraction region for each observation and each XIS detector. As the
responses of XIS0, 2 and 3 on the whole are very similar, we combined
their spectra and responses using the script \verb|addascaspec|.

We also extracted the PIN spectra. The non X-ray and cosmic X-ray
backgrounds are taken into account. The non X-ray background is
calculated from the background event files distributed by the HXD
team. The cosmic X-ray background is from the model by \citet{bo1987},
and its flux is about 5$\%$ of the background for PIN. The response
files provided by the HXD team were used. The GSO data were not used,
considering the large uncertainty in calibration and low signal to
noise ratios above 40 keV.

The CD/HID of these observations are shown in
Figure~\ref{fig:ccdiag}. We defined soft and hard colors as the ratios
of the count rates in the (3.6--5.0)/(2.2--3.6) keV bands and the
(11.0--20.0)/(5.0--8.6) keV bands, respectively. The count rates of
the lowest three energy bands were from XIS0, 1 and 3 combined. We first
obtained the count rates using the 5$\%$ pile-up exclusion regions and
then converted to the value corresponding to the whole integrated
PSF. They are background subtracted and deadtime in burst option
corrected. XIS2 was not used for this because it was not on for all
observations. The count rates in the energy band 11.0--20.0 keV were
from the PIN with the background subtracted and the deadtime
correction made. The CD and HID in Figure~\ref{fig:ccdiag} use 128-s
data. The intensity is from the energy band 2.2--8.6 keV. Two
type-I X-ray bursts were found in observation 401046010, and data
around them are not included in Figure~\ref{fig:ccdiag} or our
spectral analysis. The data points with hard color larger than 0.2
(square symbols) are all from observation 401046010, indicating that
only this observation was in the hard state while all other
observations (cross symbols) were in the soft state.

Most observations show little variation (the average and standard
deviation of colors are given in Table~\ref{tbl-1}), and we created
one spectrum each observation for spectral modeling. Each spectrum has
three instrumental components, i.e., XIS023 (combination of XIS0, 2,
3), XIS1, and PIN. In the end, we have seven spectra from {\it Suzaku}
observations, and they are denoted as $suz1$--$suz7$ hereafter
(Table~\ref{tbl-1}). We rebinned the spectra by factors of 8 and 16
for energies below and above 2.55 keV, respectively, and further
rebinning was made so that every bin has at least 40 counts and
$\chi^2$ minimization criterion can be used in our spectra fitting.

\subsection{{\it BeppoSAX} Data}

\begin{deluxetable}{lcc}
\tabletypesize{\scriptsize}
\tablecaption{The {\it BeppoSAX} observations of 4U~1705--44\label{tbl-2}}
\tablewidth{0pt}
\tablehead{Observation ID & \colhead{21292001} & \colhead{21292002}\\
  Spectral ID & \colhead{$sax1$} & \colhead{$sax2$}}
\startdata
Observation Date & 2000/08/20 & 2000/10/03\\
Exposure of LECS/MECS/PDS (ks) & 20.6/43.5/20.2 & 14.8/46.8/20.7\\
MECS2 Count Rate (cts/s)  &27.9 & 4.5\\
Spectral state & soft & hard \\
\enddata 

\end{deluxetable}

There are two pointed observations of \object{4U 1705-44} with {\it
  BeppoSAX}, one on 2000 August 20 in the soft state and the other on
2000 October 3 in the hard state
\citep[Table~\ref{tbl-2};][]{fibaub2007}. The publicly available data
are from three narrow field instruments: the Low Energy Concentrator
Spectrometer \citep[0.1--10 keV, LECS;][]{pamaba1997}, the Medium
Energy Concentrator Spectrometer \citep[1.3--10 keV,
  MECS;][]{bochco1997}, and the Phoswich Detection System
\citep[15--300 keV, PDS;][]{frcoda1997}. There are three MECS units
(MECS1, 2, 3), but no data from MECS1 are available during these two
observations. Thus we used only data from MECS2 and 3. We extracted
two spectra, one for the soft-state observation and the other for the
hard-state observation, and they are denoted as $sax1$ and $sax2$
hereafter (Table~\ref{tbl-2}). The LECS and MECS data were extracted
from circular regions of $8'$ radius centered on the source
position. As our source is in the direction of the Galactic ridge, we
cannot use ``blank fields'' measurement for background subtraction for
the LECS. Instead, we used the semi-annuli method described in
\citet{paooor1999}. For the MECS, we used the ``blank fields'' method
for the MECS as described in the instrument analysis guide. The PDS
spectra were also extracted, with the background rejection method
based on fixed Rise Time thresholds. The background for the PDS
spectra was obtained using observations during off-source
intervals. All the spectra for each instrument were finally rebinned
using the publicly available template files to sample the instrument
resolution with the same number of channels at all energies.

\section{SPECTRAL MODELING}
\label{sec:specmod}

\subsection{Spectral Models and Assumptions}
\label{sec:modeldes}

\begin{figure}
\plotone{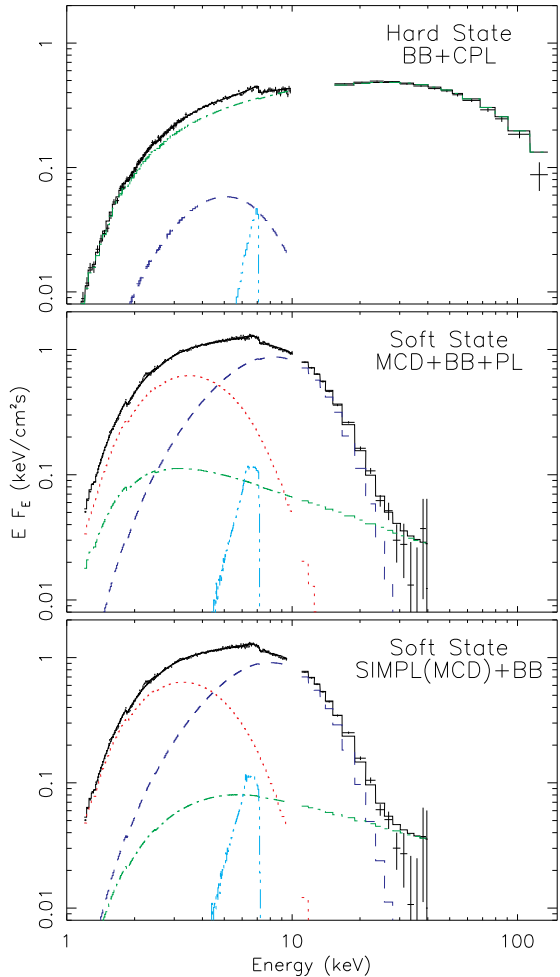}
\caption{Examples of unfolded spectra from different states using
  different models. The total model fit is shown as a black solid
  line.  The MCD component (if included) is shown by a red dotted
  line, the BB component by a blue dashed line, the PL/CPL/SIMPL
  component by a green dot-dashed line, and the Fe line (modeled by
  the diskline model) by a cyan triple-dot-dashed line. For the
  SIMPL(MCD)+BB model, the MCD component shown is the unscattered
  part, and the plotted SIMPL component refers to the scattered
  part. The hard-state spectrum is from $sax2$, and the soft-state
  spectrum from $suz2$.
\label{fig:samplefit}}
\end{figure}

We fit all nine spectra, $suz1$--$suz7$ from {\it Suzaku} and
$sax1$--$sax2$ from {\it BeppoSAX}. For {\it Suzaku}, we jointly fit
spectra from XIS023, XIS1 and the PIN. We used the energy bands with
good calibration and high signal to noise ratios for each instrument:
1.2--1.7, 1.9-2.2 and 2.3-10 keV for XIS detectors, and 11.0-40.0 keV
for the PIN. Their relative normalizations were left free, and the
spectral fit results quoted later are all from XIS023 (results from
XIS1 differ by $<3\%$ generally). For {\it BeppoSAX} spectra, we
jointly fit the LECS, MECS2, MECS3, and PDS. We utilized the 1.0--3.5
keV energy band for the LECS, 1.7--10.0 keV for the MECS, and
15.0--40.0 keV for the PDS (15.0--150.0 keV for the hard-state
spectrum $sax2$). Their relative normalizations are also left free,
but that of PDS relative to MECS was constrained to the range
0.77--0.93, a 90$\%$ confidence interval advised by the instrument
analysis guide. The spectral fit results quoted later are all
referenced to MECS2 (results from MECS3 differ by $<2\%$). The fit of
the Crab Nebula from the MECS using a single power law gives the
photon index 2.1 and normalization 9.23, while the XIS gives 2.1 and
9.55, respectively. Thus the normalizations from both observatories
appear to differ by less than 5$\%$. For all spectral fits, we set the
model systematic error to be 1$\%$.

For the soft-state spectra, we tested three models for the continuum
spectra: MCD+BB, MCD+BB+PL, and, SIMPL(MCD)+BB, respectively, where PL
is a power law and SIMPL is a simple Comptonization model by
\citet{stnamc2009}. MCD (diskbb in XSPEC) has two parameters: the
temperature $kT_{\rm MCD}$ at the apparent inner disk radius $R_{\rm
  MCD}$, and the other is the normalization $N_{\rm MCD}\equiv (R_{\rm
  MCD, km}/D_{\rm 10 kpc})^2\cos i$, where $D_{\rm 10 kpc}$ is the
distance to the source in unit of 10 kpc and $i$ is the disk
inclination. BB (bbodyrad in XSPEC) assumes an isotropic blackbody
spherical surface with radius $R_{\rm BB}$ and has two parameters,
i.e., the temperature $kT_{\rm BB}$ and the normalization $N_{\rm
  BB}\equiv (R_{\rm BB, km}/D_{\rm 10 kpc})^2$. SIMPL (in XSPEC12) is
an empirical convolution model of Comptonization in which a fraction
of the photons from an input seed spectrum is scattered into a
power-law component. This model has only two free parameters, i.e.,
the photon power-law index $\Gamma_{\rm SIMPL}$ and the scattered
fraction $f_\mathrm{SC}$. In addition, there is a flag parameter to
control whether all the scattered photons are up-scattered in energy
or are both up- and down-scattered. We specified that all the
scattered photons are up-scattered in energy. We found that inclusion
of down-scattering only changed our results within error bars. We
assumed that the Comptonization seed photons are from the disk. The
fit results of the MCD and BB are consistent within error bars if we
assumed both the MCD and BB contribute to the Comptonization seed
photons instead, thanks to low Comptonization for all our soft-state
observations. The best-fitting photon power-law index $\Gamma_{\rm
  SIMPL}$ turns out to be high ($>4$) in most cases, a regime where
the model is not suitable \citep{stnamc2009}. Thus we constrained
$\Gamma_{\rm SIMPL}$ to be less than 2.5, a value typically seen in
the black-hole cases. No constraint on the photon index $\Gamma_{\rm
  PL}$ in the PL model was used.

The two hard-state continuum spectra $sax2$ and $suz1$ were fit with a
BB plus a Comptonized component. We tested three choices of the
Comptonization component: a broken power law (BPL; bknpower in XSPEC),
a cut-off power law (CPL; cutoffpl in XSPEC), and the Comptonization
model by \citet{ti1994} (CompTT in XSPEC). CompTT is an analytic model
describing Comptonization of soft photons in a hot plasma.

All models included an absorption component (we use model wabs in
XSPEC). There are strong broad Fe lines in most spectra \citep[see
  also][]{refayo2009}. They were modeled by the diskline model
\citep{farest1989}, which describes line emission from a relativistic
accretion disk. Its parameters are: the line energy $E_{\rm line}$ in
unit of keV, the power law dependence of emissivity ($\beta$), the
disk inner and outer radii in units of $GM/c^2$, the disk inclination
$i$, and the normalization (photons cm$^{-2}$
s$^{-1}$). Figure~\ref{fig:samplefit} shows some examples of unfolded
spectra at different states using different models, with the Fe lines
modeled by the diskline model. The top panel shows the hard-state
spectrum $sax2$ using model CPL+BB. The lower two panels show the
soft-state spectrum $suz2$ using models MCD+BB+PL and SIMPL(MCD)+BB,
respectively.

We scaled the luminosity and radius related quantities using a
distance of 7.4 kpc \citep{hati1995}, unless indicated otherwise. The
flux and its error bars are all calculated for an energy band of
0.001--200 keV (1.5-200 keV for CPL/PL components) using the cflux
model in XSPEC12. As there is little emission of the thermal
components outside the above energy band, the values are essentially
bolometric for thermal components. For the MCD component, we assume
the inclination to be $24\degr$, from the fitting of Fe lines (see
below).

\subsection{Spectral Fit of the Soft-state Spectra and Results}
\label{sec:fitss}

\tabletypesize{\scriptsize}
\setlength{\tabcolsep}{0in}
\begin{deluxetable*}{lcccccccccccccr}
\tablecaption{Spectral modeling results of soft-state observations using MCD+BB+PL+diskline\label{tbl-3}}
\tablewidth{0pt}
\tablehead{ & \colhead{$kT_{\rm MCD}$ (keV)} &\colhead{$N_{\rm MCD}$} & 
  \colhead{$kT_{\rm BB}$ (keV)} &\colhead{$N_{\rm BB}$} &
  \colhead{$\Gamma_{\rm PL}$} & \colhead{$N_{\rm PL}$} &
  \colhead{$E_{\rm line}$ (keV)} & \colhead{$\beta$} &\colhead{EW (eV)}&
  \colhead{$\chi^2_\nu(\nu)$} & \colhead{$L_{\rm X, Edd}$}
}
\startdata
 suz3 & $0.82\pm 0.02$ & $114\pm 13$ & $1.87\pm 0.03$ & $6.43\pm 0.53$ & $2.71\pm 0.08$ & $0.17\pm 0.02$ & $6.87^{}_{-0.12}$ & $-4.55\pm 0.50$ & $160\pm 24$ &1.27(703)&\ $0.044\pm0.001$\\
 suz6 & $1.07\pm 0.02$ & $101\pm 6$ & $2.12\pm 0.04$ & $8.05\pm 0.57$ & $2.77\pm 0.07$ & $0.33\pm 0.02$ & $6.92^{}_{-0.08}$ & $-4.06^{+0.35}_{-0.54}$ & $176^{+11}_{-29}$ &1.05(708)&\ $0.098\pm0.002$\\
 suz2 & $1.17\pm 0.03$ & $63\pm 4$ & $2.09\pm 0.04$ & $9.53\pm 0.91$ & $2.65\pm 0.07$ & $0.30\pm 0.02$ & $6.94^{}_{-0.05}$ & $-3.64\pm 0.27$ & $160\pm 20$ &1.04(708)&\ $0.099\pm0.002$\\
 sax1 & $1.31\pm 0.03$ & $68\pm 5$ & $2.08^{+0.03}_{-0.02}$ & $14.46^{+0.83}_{-1.19}$ & $2.69^{+0.02}_{-0.04}$ & $0.62\pm 0.03$ & $6.97^{}_{-0.05}$ & $-3.59\pm 0.20$ & $156^{+41}_{-9}$ &1.08(217)&\ $0.163\pm0.003$\\
 suz5 & $1.37\pm 0.02$ & $88\pm 5$ & $2.25\pm 0.03$ & $11.73\pm 0.89$ & $2.72\pm 0.09$ & $0.38\pm 0.04$ & $6.95^{}_{-0.06}$ & $-3.45\pm 0.28$ & $154^{+36}_{-20}$ &1.00(706)&\ $0.192\pm0.004$\\
 suz7 & $1.57\pm 0.06$ & $48\pm 5$ & $2.3\pm 0.04$ & $11.5\pm 1.47$ & $2.67\pm 0.08$ & $0.44\pm 0.04$ & $6.89_{-0.08}$ & $-3.38^{+0.28}_{-0.46}$ & $165\pm 26$ &0.99(706)&\ $0.198\pm0.007$\\
 suz4 & $1.65\pm 0.07$ & $61^{+9}_{-6}$ & $2.33\pm 0.05$ & $15.51\pm 2.48$ & $2.77^{+0.16}_{-0.11}$ & $0.50\pm 0.06$ & $6.97^{}_{-0.05}$ & $-3.4\pm 0.30$ & $118^{+35}_{-17}$ &1.00(702)&\ $0.285\pm0.012$\\
\enddata 
\tablecomments{They are listed in order of the source total luminosity. See \S\ref{sec:modeldes} for the meaning of each parameter.
The normalizations of the MCD and BB models are based on the assumption of the distance to the source of 10 kpc. For the diskline model, the inclination is fixed at $24\degr$, and inner disk radius at $6GM/c^2$ (see text). EW is the equivalent width of the Fe line modeled by the diskline model, but it is not a parameter of the model. 
The last column is the total luminosity in unit of the Eddington luminosity (\S\ref{sec:fitss}), and the error bars are calculated based on simple error propagation from individual spectral components.}
\end{deluxetable*}

\tabletypesize{\scriptsize}
\begin{deluxetable*}{lcccccccccccccr}
\setlength{\tabcolsep}{0.0in}
\tablecaption{Spectral modeling results of soft-state observations using SIMPL(MCD)+BB+diskline\label{tbl-4}}
\tablewidth{0pt}
\tablehead{ & \colhead{$kT_{\rm MCD}$ (keV)} &\colhead{$N_{\rm MCD}$} & 
  \colhead{$kT_{\rm BB}$ (keV)} &\colhead{$N_{\rm BB}$} &
  \colhead{$\Gamma_{\rm SIMPL}$} & \colhead{$f_{\rm SC}$} &
  \colhead{$E_{\rm line}$ (keV)} & \colhead{$\beta$} &\colhead{EW (eV)}&
  \colhead{$\chi^2_\nu(\nu)$} & \colhead{$L_{\rm X, Edd}$}
}
\startdata
 suz3 & $0.81\pm 0.01$ & $139\pm 6$ & $1.86\pm 0.02$ & $6.67\pm 0.37$ & $2.5^{}_{-0.15}$ & $0.06\pm 0.02$ & $6.86^{}_{-0.12}$ & $-4.57^{+0.4}_{-0.59}$ & $163\pm 23$ &1.28(703)&\ $0.044\pm0.001$\\
 suz6 & $1.04\pm 0.01$ & $127\pm 4$ & $2.10\pm 0.03$ & $8.63\pm 0.59$ & $2.5^{}_{-0.16}$ & $0.04\pm 0.01$ & $6.95^{}_{-0.09}$ & $-4.28\pm 0.43$ & $173\pm 21$ &1.06(708)&\ $0.097\pm0.002$\\
 suz2 & $1.10\pm 0.02$ & $90\pm 4$ & $2.04\pm 0.04$ & $10.82\pm 0.81$ & $2.5^{}_{-0.15}$ & $0.06^{+0.01}_{-0.02}$ & $6.95^{}_{-0.06}$ & $-3.79\pm 0.25$ & $165\pm 20$ &1.05(708)&\ $0.100\pm0.002$\\
 sax1 & $1.17^{+0.02}_{-0.01}$ & $121^{+3}_{-6}$ & $2.03\pm 0.02$ & $17.16^{+1.02}_{-0.62}$ & $2.5^{}_{-0.1}$ & $0.05\pm 0.01$ & $6.97^{}_{-0.03}$ & $-3.87\pm 0.19$ & $187^{+32}_{-15}$ &1.31(217)&\ $0.165\pm0.002$\\
 suz5 & $1.39\pm 0.02$ & $88\pm 3$ & $2.27\pm 0.03$ & $11.19\pm 0.77$ & $2.5^{}_{-0.89}$ & $0.02\pm 0.01$ & $6.93^{}_{-0.06}$ & $-3.30^{+0.23}_{-0.34}$ & $161\pm 29$ &1.02(706)&\ $0.185\pm0.004$\\
 suz7 & $1.49\pm 0.03$ & $62\pm 3$ & $2.27\pm 0.04$ & $12.90\pm 1.18$ & $2.5^{}_{-0.45}$ & $0.03\pm 0.01$ & $6.9^{}_{-0.06}$ & $-3.48^{+0.25}_{-0.42}$ & $169^{+34}_{-21}$ &0.99(706)&\ $0.197\pm0.005$\\
 suz4 & $1.66\pm 0.05$ & $65\pm 5$ & $2.34\pm 0.05$ & $14.94\pm 2.06$ & $2.11^{}_{-0.61}$ & $0.01\pm 0.01$ & $6.97^{}_{-0.05}$ & $-3.37\pm 0.28$ & $117^{+34}_{-15}$ &1.00(702)&\ $0.277\pm0.010$\\

\enddata 
\tablecomments{Same as Table~\ref{tbl-3}, but for model SIMPL(MCD)+BB+diskline.}
\end{deluxetable*}

\begin{figure}
\plotone{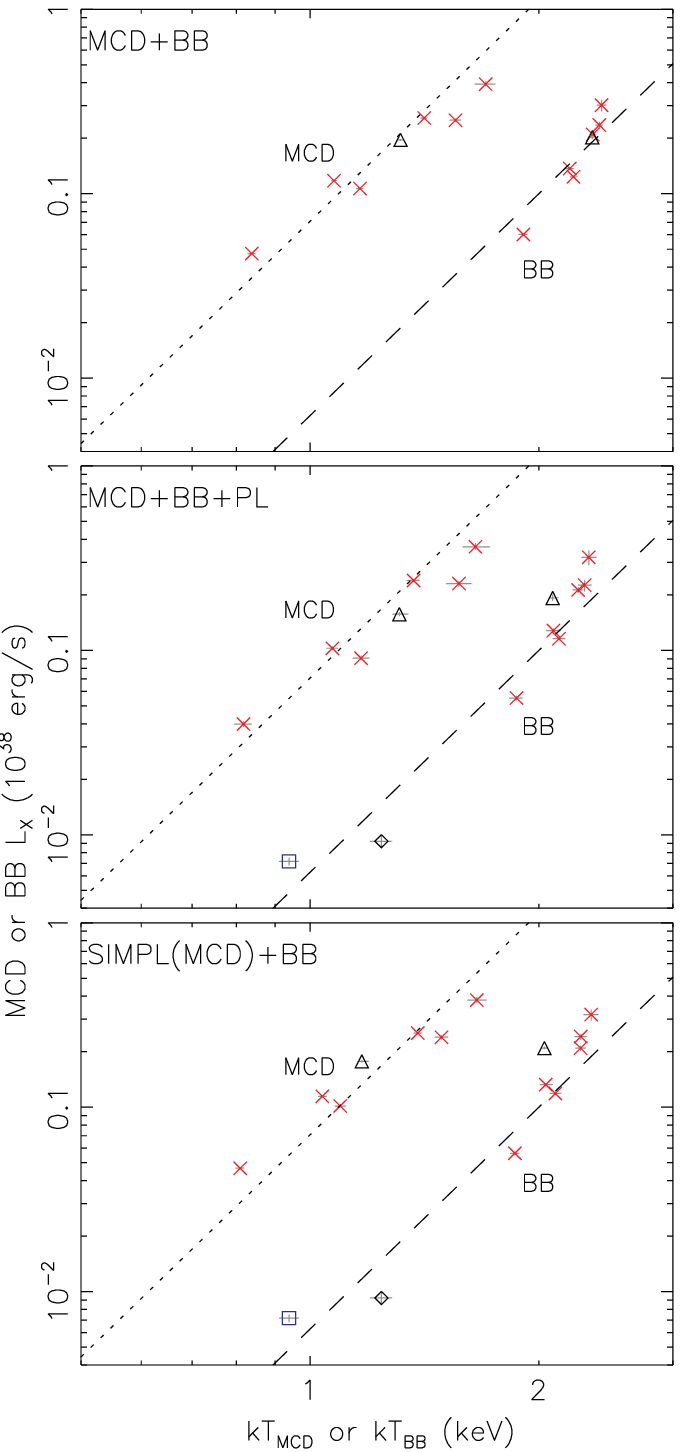}
\caption{The luminosity of the thermal components versus their
  characteristic temperatures. Each model includes the diskline
  component to fit the Fe line. The symbols in each panel denote
  different spectral states and different observatories: {\it
    BeppoSAX} hard (black diamond) and soft (black triangle) states,
  and {\it Suzaku} hard (blue square) and soft (red crosses)
  states. For the case of Model SIMPL(MCD)+BB, the MCD component shows
  the original seed spectrum luminosity (i.e., before scattering). The
  dotted lines correspond to the NS burst radius of 7.4 km (see the
  text), and the dashed lines correspond to $R_{\rm BB} =$ 2.2
  km. Error bars at a $90\%$ confidence level are also shown, but
  mostly are smaller than the symbol size.
\label{fig:spectrfitres}}
\end{figure}

We test three models for the soft-state continuum spectra: MCD+BB,
MCD+BB+PL, and SIMPL(MCD)+BB. Our practice is to use a fixed value of
the interstellar column density ($N_H$) for all spectral fits of each
model. To determine for the appropriate value, we fit all soft-state
spectra ($suz2-suz7$ and $sax1$) simultaneously with each model, while
tying their $N_{\rm H}$ to a common value. The best-fitting values of
$N_{\rm H}$ for MCD+BB, MCD+BB+PL, and SIMPL(MCD)+BB are $(1.69\pm
0.01)$, $(1.88\pm 0.06)$, and $(1.71\pm 0.01) \times 10^{22}$
cm$^{-2}$, respectively. The above values are obtained from fits with
the Fe line region 4.0--8.0 keV excluded, although the fits with the
Fe line modeled by the diskline model give very similar values. We see
that models MCD+BB and SIMPL(MCD)+BB yield similar values of $N_{\rm
  H}$, but they are smaller than that from model MCD+BB+PL. This is
consistent with the results of \citet{stnamc2009}, i.e., fits with PL
tend to infer higher values of $N_{\rm H}$ than fits with SIMPL, a
systematic effect of different ways of handling the
Comptonization. Considering this issue, we do not require a common
value of $N_{\rm H}$ for all models, but use $N_{\rm H}=1.71\times
10^{22}$ cm$^{-2}$ for models MCD+BB and SIMPL(MCD)+BB, and
$1.88\times 10^{22}$ cm$^{-2}$ for model MCD+BB+PL. In the latter
case, the choice of either value of $N_H$ does not affect the spectral
parameters for the MCD and BB components very much, but there are
obvious differences for the PL component, as will be discussed below.

The spectral fit results of the thermal components (MCD and BB) in the
soft state from all tested models with the Fe line included are shown
in Figure~\ref{fig:spectrfitres}, and results of all spectral
components are tabulated in Tables~\ref{tbl-3} (MCD+BB+PL+diskline)
and \ref{tbl-4} (SIMPL(MCD)+BB+diskline). The details of fitting the
Fe line with the diskline model are given in \S\ref{sec:feline}. The
red crosses and the black triangles in Figure~\ref{fig:spectrfitres}
are for soft-state spectra $suz2$--$suz7$ and $sax1$, respectively,
and the panels from the top to the bottom correspond to continuum
models MCD+BB, MCD+BB+PL, and SIMPL(MCD)+BB, respectively. For model
SIMPL(MCD)+BB, the MCD component shown is the original value before
scattering. The dotted lines in Figure~\ref{fig:spectrfitres}
correspond to the NS burst radius of $R_{\rm burst}$ $\sim$ 7.4 km (at
a distance of 7.4 kpc), assuming $L_{\rm X}=4\pi R^2\sigma T^4$. The
NS radius was derived from spectral fitting to Type I X-ray bursts of
this source using \xte data \citep[see also][]{gohala1989}. The dashed
lines correspond to $R =$ 2.2 km, which is about the average visible
BB emission size.

From Figure~\ref{fig:spectrfitres}, we see that the MCD and BB
components in the soft state roughly follow the $L\propto T^4$ tracks
for all models, which implies relatively constant apparent emission
areas. We discuss the extent of deviation in
\S\ref{sec:interpretation}. The inner disk radius is comparable with
the NS radius, while the visible BB emission area is about 1/11 of the
NS surface. The $kT_{\rm MCD}$ has values from $\sim$0.8 to 1.7 keV,
and $kT_{\rm BB}$ from $\sim$1.8 to 2.4 keV. These values are roughly
similar to those seen in \object{Aql X-1}, \object{4U 1608-52}, and
\object{XTE J1701-462} in the atoll soft state \citep{lireho2007,
  lireho2009}.  Fits with the Fe line region excluded from the fit
generally give consistent results for the MCD and BB components, to
within 10\% for the normalization parameter $N_{\rm MCD}$ and within
20\% for the $N_{\rm BB}$. Spectrum $sax1$ gives larger differences
($\sim 40\%$ for $N_{\rm BB}$), probably because of its narrower
energy band and lower energy resolution (only five channels above 10
keV and none in the energy band of 10--15 keV).

Results of the thermal components from both models MCD+BB+PL and
SIMPL(MCD)+BB are generally similar, but differences of $>20\%$ can
occur in some cases (e.g., 30\% for $N_{\rm MCD}$ from the spectrum
$suz2$; compare Tables~\ref{tbl-3}--\ref{tbl-4}). Due to weak
Comptonization in the soft state, model MCD+BB in general also gives
similar results of the thermal components (e.g., flux differs $<5\%$
and $N_{\rm BB}$ by $<20\%$) and acceptable reduced $\chi^2$ values
($<2.0$; but 2.6 for $sax1$). The largest differences in best-fitting
spectral parameters from model MCD+BB compared with those from models
MCD+BB+PL and SIMPL(MCD)+BB are from {\it BeppoSAX} spectrum $sax1$
($N_{\rm BB}$ differs by $\sim$50\%).

\subsection{Spectral Fit of the Hard-state Spectra and Results}
\label{sec:fiths}

\begin{deluxetable*}{lcccccccccccccc}
\addtolength{\tabcolsep}{-5pt}
\tabletypesize{\scriptsize}
\tablecaption{Spectral modeling results of hard-state observations using BB+CPL+diskline\label{tbl-5}}
\tablewidth{0pt}
\tablehead{ & \colhead{$N_{\rm H}$ ($10^{22}$ cm$^{-2}$)} &
  \colhead{$T_{\rm BB}$ (keV)} &\colhead{$N_{\rm BB}$} &
  \colhead{$\Gamma_{\rm CPL}$} & \colhead{$N_{\rm CPL}$} &
  \colhead{$E_{\rm line}$ (keV)} & \colhead{$\beta$} &\colhead{EW (eV)}&
  \colhead{$\chi^2_\nu(\nu)$} & \colhead{$L_{\rm X, Edd}$}
}
\startdata
suz1 & 1.71f & $0.94\pm 0.02$ & $17.62\pm 1.58$ & $1.19\pm 0.02$ & $0.056\pm0.002$ & $6.60\pm 0.04$ & $-2.45\pm 0.28$ & $92\pm 27$ &1.10(676)&$0.055\pm0.001$\\
sax2 & 1.71f & $1.30\pm 0.03$ & $6.06\pm 0.61$ & $1.37\pm 0.01$ & $0.119\pm0.003$ & $6.80^{}_{-0.13}$ & $-2.46\pm 0.55$ & $93\pm 36$ &1.19(228)&$0.070\pm0.001$\\
\\
suz1 & 1.88f  & $0.94\pm 0.03$ & $13.12\pm 1.62$ & $1.33\pm 0.01$ & $0.076\pm0.002$ & $6.59\pm 0.04$ & $-2.30\pm 0.33$ & $74\pm 28$ &1.06(676)&$0.049\pm0.001$\\
sax2 & 1.88f  & $1.24\pm 0.04$ & $5.52\pm 0.77$ & $1.40\pm 0.01$ & $0.135\pm0.003$ & $6.87^{}_{-0.15}$ & $-2.86\pm 0.55$ & $126\pm 37$ &0.93(228)&$0.070\pm0.001$\\

\enddata 
\tablecomments{Same as Table~\ref{tbl-3}, but for model BB+CPL+diskline and for the hard-state data. The cutoff energy of the CPL is fixed at 44 keV.}
\end{deluxetable*}

\begin{figure}
\plotone{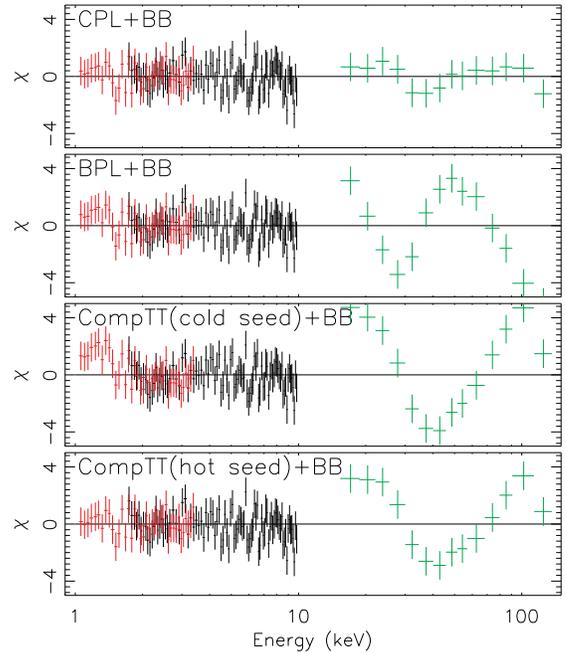} 
\caption{The fit residuals in terms of sigmas with error bar values
  set to be one for different Comptonization models in combination
  with BB model, using the hard-state spectrum $sax2$. The red, black,
  green points are from LECS, MECS2, and PDS, respectively.
\label{fig:comparecompmodel}}
\end{figure}

\begin{figure}
\plotone{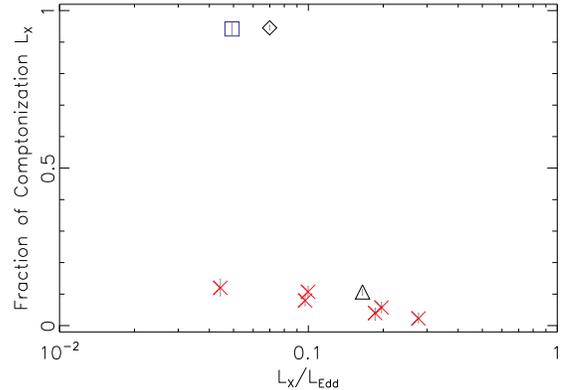} 
\caption{The energy fraction of Comptonized luminosity versus the
  total luminosity, from model SIMPL(MCD)+BB (the soft state) and
  model CPL+BB (the hard state). See Figure~\ref{fig:spectrfitres} for
  meanings of symbols. Error bars at a $90\%$ confidence level are
  shown for the fraction of Comptonization Lx, and those for Lx/LEdd
  are not shown, but all are very small
  (Tables~\ref{tbl-3}--\ref{tbl-5}).
\label{fig:nontherfrac_lum}}
\end{figure}

\begin{figure}
\plotone{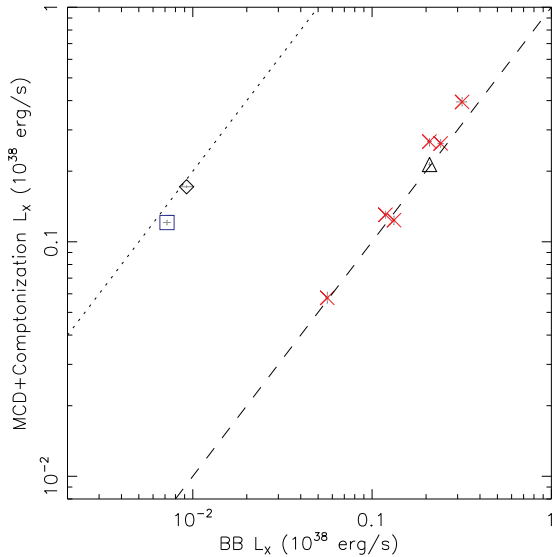} 
\caption{The luminosity of the MCD component plus Comptonization
  versus the BB luminosity. The results for the soft state are from
  model SIMPL(MCD)+BB+diskline, and the Comptonization refers to the
  scattered disk emission, modeled by SIMPL. Model MCD+BB+PL+diskline
  gives similar results. The dashed and dotted lines mark the ratios
  of 1 and 0.05 of the BB luminosity versus the MCD component plus
  Comptonization luminosity, respectively. See
  Figure~\ref{fig:spectrfitres} for meanings of symbols. Error bars at
  a $90\%$ confidence level are also shown, but mostly are smaller
  than the symbol size.
\label{fig:bpldisk_bbd}}
\end{figure} 

We first investigate the performance of different Comptonization
models. Using spectrum $sax2$, which extends to 150 keV (instead of
$40$ keV for spectrum $suz1$), we obtained values of $\chi^2/\nu$
190.2/226, 287.1/224, and 339.9/225 for models CPL+BB, BPL+BB, and
CompTT+BB, respectively ($N_{\rm H}$ was allowed to have different
values for different models). The result for CompTT+BB quoted above is
just one possible solution with a local $\chi^2$ minimum. The
best-fitting input seed photon temperature $\tau_0$ is $<$0.2 keV
\citep[the cold seed photon model;][]{lireho2007}. All the above
models give similar results in terms of $kT_{\rm BB}$ ($\sim 1.5$
keV), $N_{\rm BB}$ ($\lesssim 20$), and the Comptonization fractions
($>90\%$). There is another solution for CompTT+BB, which has
$\chi^2/\nu=252.5/225$. It has $\tau_0=0.9\pm 0.1$ keV \citep[the hot
  seed photon model;][]{lireho2007}. For this case, the inferred BB
component is quite different from the models above, with $kT_{\rm
  BB}=0.5\pm 0.1$ keV and $N_{\rm BB}=256\pm110$. The fit residuals of
all the above models are shown in
Figure~\ref{fig:comparecompmodel}. We see that CPL+BB gives the best
fit over the whole energy range, while other models tend to have large
residuals at high energy ($>15$ keV). Hereafter we focus on model
CPL+BB only.

The fit of the CPL model to the {\it BeppoSAX} hard-state 15.0--150.0
keV spectrum (the BB contributes little at energies above 15.0 keV)
gives a cutoff energy of $44\pm4$ keV. The cutoff energy cannot be
well constrained in the case of spectrum $suz1$, for which we use
energies only up to $40$ keV. Thus we fixed the cutoff energy to be 44
keV when we fit spectra $sax2$ and $suz1$. The simultaneous fit of
$sax2$ and $suz1$ gives a best-fitting value of $N_{\rm H}=(1.89\pm
0.04)\times 10^{22}$ cm$^{-2}$ from model CPL+BB, close to the value
of $1.88\times 10^{22}$ cm$^{-2}$ obtained from model MCD+BB+PL for
the soft state. For comparison, we fit the hard-state spectra using
both $N_{\rm H}=1.88$ and $1.71\times 10^{22}$ cm$^{-2}$, and the
results are given in Table~\ref{tbl-5}. We see no large different
between these two sets of solutions. Using $N_{\rm H}=1.88\times
10^{22}$ cm$^{-2}$ tends to give a smaller BB emission area and a
higher photon index of the CPL model. The values of the photon index
are more similar between spectra $sax2$ and $suz1$ than using $N_{\rm
  H}=1.71\times 10^{22}$ cm$^{-2}$. Both values of $N_{\rm H}$ give
high fractions of Comptonization ($>$90$\%$).

The results using $N_{\rm H}=1.88\times 10^{22}$ cm$^{-2}$ are shown
in Figure~\ref{fig:spectrfitres}, the blue squares and the black
diamonds for $suz1$ and $sax2$, respectively. They are repeated in the
middle and bottom panels. They are not shown in the top panel, because
this panel is reserved for solutions free of Comptonization and such a
model is unacceptable for observations of the hard state. From
Figure~\ref{fig:spectrfitres}, we see that the boundary layer emission
areas in the hard state are comparable to those in the soft
state. Although $L_{\rm BB}$ changes over 50 times, the values of
$N_{\rm BB}$ are always in the range of $\sim$5-15.

One of our main goals of using the broad-band spectra is to search for
detectable thermal emission from the accretion disk in the hard
state. We added the MCD component while analyzing the two hard-state
spectra ($suz1$ and $sax2$), i.e., using model CPL+MCD+BB, with either
the Fe modeled with the diskline model or with the Fe region excluded
in the fit. $N_{\rm H}$ is either fixed at 1.71 or 1.88 $\times
10^{22}$ cm$^{-2}$, or left free in the fit. The best-fitting disk
temperature tends to go below 0.2 keV, with an upper limit $<0.25$ at
a 90\%-confidence level. The normalizations of the MCD component
$N_{\rm MCD}$ are not well constrained. For all cases, the unscattered
flux of the MCD component is $<2\%$ of the total flux (absorbed or
unabsorbed; 1--200 keV). Thus, if we assume a physically visible disk,
then we cannot exclude either possibility, i.e., that the disk in the
hard state might be truncated at a very large radius and/or the
temperature is below 0.2 keV. Alternatively, the disk may be rendered
invisible by very high Comptonization.

Figure~\ref{fig:nontherfrac_lum} shows the fraction of Comptonized
luminosity versus the total luminosity, using SIMPL(MCD)+BB for the
soft state and CPL+BB for the hard state. The total luminosity is
normalized by the Eddington luminosity $L_\mathrm{Edd}$, which is
derived from the type I X-ray bursts showing photospheric radial
expansion and corresponds to an average peak flux of about
$4\times10^{-8}$ \ergcm
\citep{gapsmu2006}. Figure~\ref{fig:nontherfrac_lum} shows that
Comptonization only constitutes $<15\%$ of the emission in the soft
state, but $>90\%$ in the hard state. For the soft-state data, the
fractional contribution of Comptonized luminosity decreases with
luminosity on the whole, which is consistent with the behavior of
atoll-type transients \citep{lireho2007}. The MCD+BB+PL model also
gives low-Comptonization solutions for the soft state ($<15\%$), with
$N_{\rm H}$ either 1.71 or 1.88 $\times 10^{22}$ cm$^{-2}$. One main
difference between these two choices of $N_{\rm H}$ is that fits with
$N_{\rm H}=1.88\times 10^{22}$ cm$^{-2}$ give higher values of
$\Gamma_{\rm PL}$ ($\sim 2.7$) than fits with $N_{\rm H}=1.71\times
10^{22}$ cm$^{-2}$ ($\Gamma_{\rm PL}\sim 2.2$).

The BB component, which describes the emission from the boundary layer
in our model, is present in both the hard and soft states. We can
compare this component with the other spectral components to
investigate the impact of different accretion processes in different
states. We plot in Figure~\ref{fig:bpldisk_bbd} the luminosity of the
MCD component plus Comptonization (SIMPL/CPL) versus the BB luminosity
from model SIMPL(MCD)+BB. Model MCD+BB+PL gives similar results. The
hard-state data (diamond and square symbols) are from model CPL+BB.
The dashed and dotted lines correspond to the ratios of 1 and 0.05 of
the BB luminosity versus the MCD component plus Comptonization
luminosity, respectively. They are about the average values for the
soft- and hard-state observations, respectively. If we assume that the
Comptonization emission is not from the boundary layer, the above
result means that there is a much lower portion of energy seen in the
visible portion of the boundary layer in the hard state than in the
soft state. Similar results were suggested to be a possible
consequence of a strong jet in the hard state in \citet{lireho2007}.

\subsection{Relativistic Fe lines}
\label{sec:feline}

\begin{figure}
\plotone{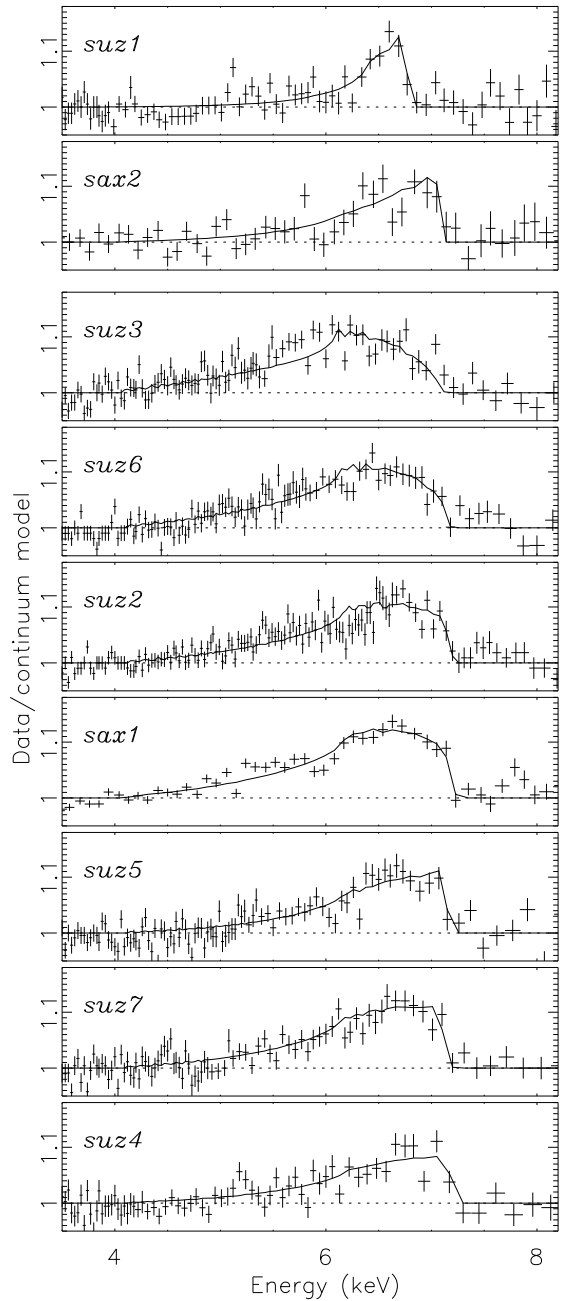} 
\caption{Fe lines of all spectra, fit by the diskline model. The upper
  two panels are for the hard state (spectra $suz1$ and $sax2$), while
  others for the soft state. The soft-state continuum is fit by model
  SIMPL(MCD)+BB, while model MCD+BB+PL gives very similar
  results. From top to bottom, the source luminosity increases, except
  for the hard-state spectra $suz1$ and $sax2$, which are put on the
  top panels and have luminosities between that of $suz3$ and $suz6$.
\label{fig:ironline}}
\end{figure} 

\begin{figure}
\plotone{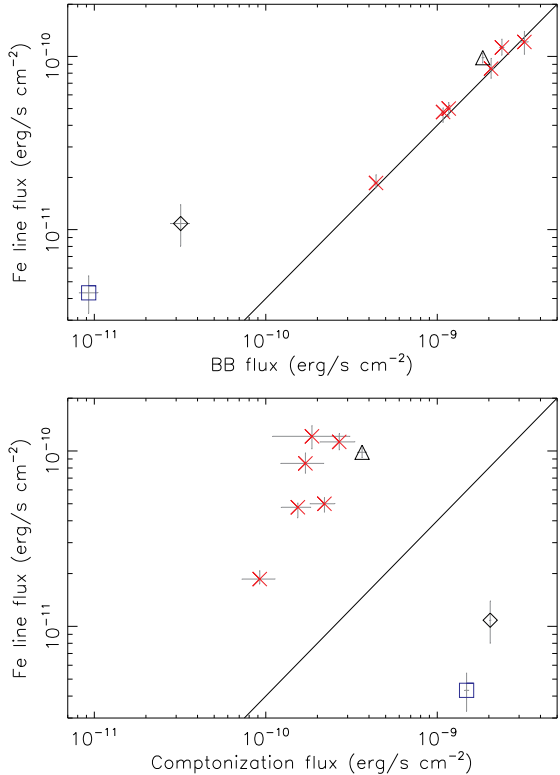}
\caption{Dependence of Fe line flux on the BB flux (upper panel) and
  Comptonization flux (SIMPL/CPL components; lower panel). The solid
  lines mark the ratio of $4\%$ of the Fe line flux over the BB flux
  or the Comptonization flux. The Fe line flux is bolometric, but the
  BB flux and Comptonization flux are integrated over 6.4--200
  keV. See Figure~\ref{fig:spectrfitres} for meanings of
  symbols. Error bars at a $90\%$ confidence level are also shown, but
  mostly are smaller than the symbol size.
\label{fig:ironlineflux}}
\end{figure}

We first fit the Fe line with a Gaussian line. A broad Gaussian line
with the line width $\sigma$ around 0.6 is generally required, while
fits with a narrow line ($\sigma<0.1$ keV) are mostly unacceptable,
with $\chi^2$ values increased by $>100$ for $\sim$700 degrees of
freedom in the soft state. The diskline model can in general improve
the fits further, compared with a broad Gaussian line, with the
$\chi^2$ values decreased by about 30 on average in the soft state. In
the hard state, the diskline model still gives the best fits, though
the improvement is much less. We also fit the spectra with a smeared
edge (smedge in XSPEC) in our continuum model, with the lower limit of
the threshold energy set to be the neutral Fe K edge 7.1 keV. We
obtain fits with $\chi^2$ values larger than those using the diskline
model by about 30 on average, and the threshold energy in most fits
reaches the lower limit 7.1 keV. Thus the broad line feature in the
spectra is probably not completely due to the Fe absorption edge. The
more complicated model combining both a Gaussian line and a smeared
edge can give fits with similar $\chi^2$ values as those using the
diskline model, but we choose to focus on the results using the
simpler model, i.e., the diskline model.

While fitting the Fe line with the diskline models, we initially fit
all soft-state spectra simultaneously, with the inclination parameter
$i$ tied to a common value. Different continuum models turn out to
give quite similar best-fitting values of $i$: $24.3\pm 0.8\degr$,
$23.3\pm 1.0\degr$, and $24.2\pm 0.8\degr$ from models MCD+BB,
MCD+BB+PL, and SIMPL(MCD)+BB, respectively. Thus we fix the disk
inclination to be $i=24\degr$. For most of the soft-state spectra, the
inner disk radius inferred from the diskline model reaches 6 $GM/c^2$,
the innermost stable circular orbit (ISCO). Thus in the final fitting,
we fix it to be 6 $GM/c^2$. For the hard-state spectra, the Fe lines
are much weaker, and the best-fitting inner disk radius in the
diskline model is quite uncertain. The lower error bar reaches 6
$GM/c^2$ for the two observatories using both $N_{\rm H}=1.88$ and
$1.71\times 10^{22}$ cm$^{-2}$. For simplicity, we then fixed the
inner disk radius in the diskline model to be 6 $GM/c^2$ for the
hard-state spectra to derive the final results. The obtained Fe line
flux and equivalent width are not sensitive to such details.

The results of fitting Fe lines with the diskline model are shown in
Figure~\ref{fig:ironline}. It can be seen that the Fe emission line is
detected in all spectra, and most of them show a broad feature
\citep[see also][]{refayo2009}. Tables~\ref{tbl-3}--\ref{tbl-5}
contain best-fitting parameters of the Fe line for each spectrum, with
the equivalent width also included. The best-fitting value of $E_{\rm
  line}$ reaches the allowed upper limit 6.97 keV in most cases. The
equivalent width is around 170 eV for most of the soft-state spectra,
and it is around 100 eV in the hard state. The power-law dependence of
emissivity $\beta$ is generally steeper in the soft state ($\sim-3.5$)
than in the hard state ($\sim-2.5$).

Fe emission lines in X-ray binaries are the most obvious signature of
an accretion disk irradiated by an external source of hard X-rays, due
to a combination of high fluorescent yield and large cosmic abundance
\citep{mi2007}. To investigate the irradiation source of the Fe lines,
we show in Figure~\ref{fig:ironlineflux} the dependence of the
bolometric Fe energy line flux on the BB energy flux (upper panel) and
Comptonization (SIMPL/CPL) flux (lower panel), both integrated over
6.4--200 keV. The soft-state data show that the Fe line flux increases
monotonically with the BB flux, but has no clear dependence on the
Comptonization flux. This might imply that it is the boundary layer
emission that illuminates the accretion disk and produces the Fe line,
in agreement with the conclusions of \citet{camiba2009}. The Fe line
energy flux is about $4\%$ of the BB energy flux (solid lines).

The hard-state data in Figure~\ref{fig:ironlineflux} (diamond/square
symbols) in the upper panel show that there is much higher Fe line
flux relative to the BB continuum in the hard state compared to the
soft state. This might indicate that the BB emission is not the
(major) source illuminating the accretion disk to produce the Fe line
in the hard state. From the lower panel, we can see that there is
strong Comptonization flux available in the hard state. However, using
the reference lines and comparing the upper and lower panels, we see
that the Comptonization emission is not as efficient as the BB
emission to illuminate the accretion disk and produce the Fe
line. This can be explained if the Comptonization emission is located
farther away from the accretion disk than the boundary layer. We also
caution that some BB flux may be screened from the observer's view by
the accretion disk.

\section{DISCUSSION}
\label{sec:interpretation}
\subsection{Global View of Two Spectral States}
We have shown that \object{4U 1705-44} exhibits two distinct spectral
states. The hard-state spectra span from $\sim$1 to 150 keV, while the
soft-state spectra are mostly confined below 40 keV \citep[see
  also][]{fibaub2007}. Such two distinct states have been observed in
other atoll-type NS LMXBs \citep{ba2001,va2006}. On the whole, these
two spectral states are very similar to the thermal and hard states
(but not the steep power law) of accreting black holes
\citep{remc2006,dogiku2007}. In addition, atoll-type NS LMXBs and
accreting black holes also have many similarities in timing
properties. All this motivates the speculation that the accretion
process is quite similar between atoll-type NS LMXBs and accreting
black holes \citep{ba2001,dogiku2007,lireho2007}. This implies that
the presence of the boundary layer in atoll-type NS LMXBs probably
does not strongly interfere with the accretion disk, at least at
luminosities below $\sim$0.5 $L_{\rm Edd}$. However, the boundary
layer is still an important spectral component in atoll-type NS
LMXBs. Figure~\ref{fig:bpldisk_bbd} shows that the boundary layer
emission flux is comparable with that from the disk. Because the
boundary layer is hotter than the disk \citep{miinko1984}, the
soft-state spectra tend to cover a broader energy band and extend to
higher energies in atoll-type NS LMXBs than in the thermal state of
accreting black holes (compare our Figure~\ref{fig:samplefit} and
Figure~2 in \citet{remc2006}). The relatively weak contribution of the
boundary layer in the hard state, as shown in
Figure~\ref{fig:bpldisk_bbd}, can be interpreted as the consequence of
mass ejection in this state \citep{lireho2007}. This seems to be in
alignment with the jet model for the hard state in accreting black
holes \citep{fe2006,dogiku2007}.

\subsection{Inner Disk Radius}

In \citet{lireho2007}, our spectral modeling of two atoll sources
\object{Aql X-1} and \object{4U 1608-52}, using the MCD+BB model plus
weak Comptonization showed the MCD behavior close to the $L_{\rm
  MCD}\propto T_{\rm MCD}^4$ track. In contrast, classical
two-component models resulted in constant or slightly decreasing
temperatures of the thermal component (MCD or BB) with increasing
luminosity. We note, however, that in that study we used \xte data and
the MCD component in the low-luminosity soft state could not be well
constrained. In this study, we used {\it Suzaku} and {\it BeppoSAX}
spectra, which extend the energy range down to 1 keV, and the MCD
components are well constrained for all soft-state spectra. As the
result of the better low-energy coverage, we did not need to put
additional constraints on the PL component, as was necessary in
\citet{lireho2007}. This motivates us to evaluate more quantitatively
how closely the MCD component follows the $L_{\rm MCD}\propto T_{\rm
  MCD}^4$ track.

Our spectral modeling of broad-band spectra of \object{4U 1705-44}
shows a track that is flatter than the $L_{\rm MCD}\propto T_{\rm
  MCD}^4$ relation, with nearly an order of magnitude variation in
luminosity. There is a slight decrease of $R_{\rm MCD}$ at higher
luminosity. Spectral fitting with Comptonization modeled by PL
suggests $L_{\rm MCD}\propto T_{\rm MCD}^{3.3\pm0.2}$, while fitting
by SIMPL results in $L_{\rm MCD}\propto T_{\rm MCD}^{3.1\pm
  0.1}$. Such deviations have been seen in several black-hole X-ray
binaries, and it is normally believed to be due to a spectral
hardening effect, instead of a real change in the inner disk radius
\citep{shmcna2006, dadobl2006, mcnash2007, mcreru2009}. Spectral
hardening arises when the electron scattering dominates over
absorption as the main opacity source. In such a situation, the local
specific flux in the disk appears as a simple dilute blackbody with a
color temperature higher than the effective temperature by a factor of
$f_{\rm col}$ \citep{shta1995}. This factor slightly increases with
luminosity/temperature. Based on a simple analytic estimate of the
hardening factor, \citet{dadobl2006} suggested a $L_{\rm MCD}\propto
T_{\rm MCD}^{3}$ relation. It should be noted that the extent to which
the hardening factor depends on luminosity can vary with the
inclination, the mass of the compact object, etc. A detailed numeric
simulation to obtain how the hardening factor behaves for an accreting
NS or direct accretion modeling to infer the real inner disk radius
with more realistic spectral models incorporating many effects could
provide further insights. We further note that there are possibly
other factors causing the above deviation. We cannot exclude the
possibility of real change of the inner disk radius in some part of
the soft state. It is also possible that the above deviation is due to
our simple descriptions of the weak Comptonization and the boundary
layer.

\subsection{Constraint on the Magnetic Field in 4U~1705--44}
We perform a rough estimate of the magnetic field in \object{4U
1705-44}, under the assumption that the disk is truncated at the ISCO
in our soft-state observations. This requires the magnetic field to be
dynamically unimportant for these soft-state observations. That is,
the Alfv\'{e}n radius $r_{\rm A}$, the radius at which the magnetic
pressure is roughly the sum of the ram and gas pressure, should be
smaller than the ISCO. Based on Equations 6.19--6.20 in
\citet{frkira1985}, we have
\begin{eqnarray}
r_{\rm A}\sim & 7.5 \left(\frac{k_{\rm A}}{0.5}\right) \left(\frac{M_{\rm NS}}{M_{\odot}}\right)^{1/7} \left(\frac{R_{\rm NS}}{10\ {\rm km}}\right)^{10/7}  \nonumber\\
            &  \left(\frac{L}{10^{37}\ {\rm erg/s}}\right)^{-2/7} \left(\frac{B}{10^{8}\ {\rm G}}\right)^{4/7}\ {\rm km},
\end{eqnarray}
where $B$ is the magnetic field strength at the surface of the NS and
$k_{\rm A}$ is the correction from the spherical accretion to disk
accretion and is about 0.5. This formula assumes a dipole magnetic
field. We further assume $M_{\rm NS}=1.4$ \msun\ (such that $R_{\rm
  in}=12.4$ km), and $R_{\rm NS}=10$ km. The above expression shows
that the obtained value of $B$ will only weakly depend on the NS mass
(a power of $1/4$). Using $L=10^{37}$ erg/s, from the faintest
soft-state spectrum $suz3$, and the constraint $r_{\rm A}<R_{\rm in}$,
we find $B<1.9\times 10^8$ G. We can also assume that $r_{\rm A}$ is
less than $R_{\rm NS}$, which would decrease the above limit by
30$\%$.

\subsection{The Boundary Layer}
We see that the apparent emission of the boundary layer, modeled by
BB, roughly follows $L_{\rm BB}\propto T_{\rm BB}^{4}$, from the hard
to the soft states with the luminosity of the boundary layer covering
$\sim$0.003-0.12 $L_{\rm Edd}$. The apparent area of the boundary
layer is about 1/11 of the the NS surface, as inferred from Type I
X-ray bursts. The actual area of the boundary layer might be larger
due to a special geometry of the boundary layer, which has been
attributed to be an equatorial belt \citep{lireho2007}. Using Equation
3 in \citet{lireho2007} and using the inclination obtained from the Fe
line fit ($i=24\degr$), the latitude range (from the NS equator) of
the boundary layer is about $13\degr$, corresponding to an emission
area of $23\%$ of the NS surface. Accordingly $L_{\rm BB}$ should be
1.6 times larger than shown in \S~\ref{sec:specmod}, which assumes
isotropic emission of the boundary layer. Thus, $L_{\rm BB}/L_{\rm
  MCD+Comptonization}$ is not around 1 in the soft state, as shown in
Figure~\ref{fig:bpldisk_bbd}, but is about 2.6. If $i=60\degr$, the
latitude range of the boundary layer is about $7\degr$, corresponding
to an emission area of $12\%$ of the NS surface. In this case $L_{\rm
  BB}$ should be $30\%$ larger than shown in
\S~\ref{sec:specmod}. $L_{\rm MCD}$ should increase by $80\%$ (i.e.,
adjusting for $i=60\degr$) so that $L_{\rm BB}/L_{\rm
  MCD+Comptonization}$ would be around 0.7 in the soft state.

Whether the above results imply that the real boundary layer area is
small and nearly constant further depends on the radiative transfer
process in the atmosphere above the boundary layer, i.e., the
hardening effect as discussed above for the disk spectra. The small BB
emission area is reminiscent of the well known spectral modeling
problem of the NS thermal emission in quiescence, i.e., the BB fit of
its thermal component produced inferred radii too small for
theoretical NS size estimate, whereas models taking into account the
radiative transfer in the hydrogen atmosphere give radius estimate
much closer to theoretical expectation of the size of NSs
\citep[e.g.,][]{rubibr1999}. However, all our observations are quite
bright ($\gtrsim 10^{37}$ erg/s), and the emission should be due to
active accretion. At such a high accretion rate, a pure hydrogen
atmosphere is not expected \citep{brbiru1998}, and the above problem
might not apply to our case. Our conclusion of the small size of the
boundary layer is based on the assumption that the modification of the
bursting atmosphere on burst emergent spectra is similar to that of
the boundary layer emission. This assumption might be valid if most of
the heat in the boundary layer is generated in a layer as deep as that
for burst nuclear burning. We note that the small inferred size of the
boundary layer agrees with the theoretical expectation of most of the
boundary layer models at sub-Eddington accretion rates
\citep{klwi1991,insu1999,posu2001}. Thus it is quite possible that the
boundary layer emission area is indeed small for our observations.

If the behavior of the hardening factor for the boundary layer is
similar to that for the burst emission, then one might conclude that
the boundary layer emission area is constant if $L_{\rm BB}\propto
T_{\rm BB}^{4}$ is measured, as the hardening factor is quite
independent of the temperature for burst emission in sub-Eddington
limit \citep{majoro2004,oz2006}. For example, for a NS with mass 1.4
\msun and radius 10 km, Table 2 from \citet{majoro2004} suggests an
approximate relation of $L_{\rm BB}\propto T_{\rm BB}^{3.7}$ for
$T_{\rm BB}$ within 1.1--2.5 keV (most of our observations fall into
this range). Thus, our results of roughly following $L_{\rm BB}\propto
T_{\rm BB}^{4}$ from the hard to soft states might imply little change
of the real boundary layer emission area for our observations.

We note that there is scatter in the inferred apparent emission area
of the boundary layer. The fractional variation is about $40\%$,
larger than the typical error bars (10$\%$). There are also systematic
differences in the inferred boundary layer emission area between {\it
  BeppoSAX} and {\it Suzaku}, both in the hard and soft states (See
Tables~\ref{tbl-3}--\ref{tbl-5}). Whether all this is real or affected
by systematic problems related to spectral models and/or
instrumentation is unclear. If we do a simple power-law fit as we did
for the MCD component, the BB component seems to follow $L_{\rm
  BB}\propto T_{\rm BB}^{5.0 \pm 0.2}$ from fit to all data combined
and $L_{\rm BB}\propto T_{\rm BB}^{4.3 \pm 0.1}$ from fit to {\it
  Suzaku} data only. Because of the above uncertainties, we did not
treat these deviations from $L_{\rm BB}\propto T_{\rm BB}^{4}$ as
significant.  A better understanding of our luminosity vs. temperature
results for the disk and boundary layer of an accreting NS may be
gained from further theoretical work on each component, and from
additional observations, e.g., with improved statistics and dynamic
range for samples of the hard state.

\section{CONCLUSION}
\label{sec:conclusion}
The broad-band X-ray spectra of \object{4U 1705-44} obtained with {\it
  Suzaku} and {\it BeppoSAX} show two distinct hard and soft spectral
states. These spectra have significantly better coverage in the soft
X-ray energy band compared with those from \xte. We have successfully fit
these spectra using the model similar to \citet{lireho2007}. 

The accretion disk in the soft state seems to approximately follow a
$L\propto T^{3.2}$ track. One cause of the deviation from $L\propto
T^4$ maybe a luminosity-dependent spectral hardening factor. However,
it is still possible that the inner disk radius is really changing in
some part of the soft state. We found no significant contribution of
the thermal disk in our hard-state spectra above 1 keV, and the disk
might be truncated at a large radius and/or has a low temperature
($<0.2$ keV), or is buried under high Comptonization. The boundary
layer is roughly constant from the hard to soft states, with apparent
emission size about 1/11 of the whole surface of the neutron
star. Assuming that the disk is truncated by the ISCO in the soft
state or the NS surface, we estimated the magnetic field of in
\object{4U 1705-44} to be less than about $1.9\times 10^8$ G.

Broad relativistic Fe lines are also detected in most of the spectra,
especially in the soft state. We modeled them with the diskline model
and found that the strength of the Fe line correlates well with the
boundary layer emission in the soft state, with the Fe line flux about
4\% of the flux from the boundary layer ($>6.4$ keV). In the hard
state, our results suggest that the Fe lines are due to the strong
Comptonization emission. However, the Comptonization emission in the
hard state seems to illuminate the accretion disk and produce the Fe
line not as efficiently as the boundary layer emission in the soft
state, probably because the boundary layer is closer to the inner
accretion disk.

The authors would like to express their thanks to all members of the
Suzaku team members, especially Koji Mukai, for their support in the
schedule of observations and preparation of this paper. Support for
this research was provided by the NASA Grant NNX08AC02G under the
Suzaku guest observer program and the NASA contract to MIT for RXTE
instruments. D.L. thanks

\end{document}